\documentclass[12pt]{article}
\usepackage[margin=2cm]{geometry}
\usepackage{amsmath,amssymb,mathtools,graphicx}
\usepackage{cite}
\usepackage{slashed}
\usepackage{color}
\makeatother

\newcommand{\be}{\begin{equation}}
\newcommand{\bea}{\begin{eqnarray}}
\newcommand{\eea}{\end{eqnarray}}
\newcommand{\ba}{\begin{array}}
\newcommand{\ea}{\end{array}}
\newcommand{\ee}{\end{equation}}
\newcommand{\bes}{\begin{equation*}}
\newcommand{\beas}{\begin{eqnarray*}}
\newcommand{\eeas}{\end{eqnarray*}}
\newcommand{\bas}{\begin{array*}}
\newcommand{\eas}{\end{array*}}
\newcommand{\ees}{\end{equation*}}

\numberwithin{equation}{section}
\begin{document}

\begin{titlepage}
\vspace{10mm}
\begin{flushright}
\end{flushright}

\vspace*{20mm}
\begin{center}

{\Large {\bf  Island in the Presence  of Higher Derivative Terms }\\
}

\vspace*{15mm}
\vspace*{1mm}
{Mohsen Alishahiha${}^a$,   Amin Faraji Astaneh$^{b,c}$ and  Ali Naseh${}^c$ }

 \vspace*{1cm}

{\it ${}^a$ School of Physics,
Institute for Research in Fundamental Sciences (IPM)\\
P.O. Box 19395-5531, Tehran, Iran
\\ \vspace{0.3cm}
 ${}^b$Department of Physics, Sharif University of Technology, Tehran 14588-89694, Iran
 \\ \vspace{0.3cm}
 ${}^c$  School of Particles and Accelerators, Institute for Research in Fundamental Sciences (IPM)\\
P.O. Box 19395-5531, Tehran, Iran
  }

 \vspace*{0.5cm}
{E-mails: {\tt alishah@ipm.ir, faraji@sharif.ir, naseh@ipm.ir}}%

\vspace*{1cm}
\end{center}

\begin{abstract}
Using extended island formula we compute  entanglement entropy of Hawking radiation
for black hole solutions of certain gravitational models containing higher derivative terms.
To be concrete we consider two different four dimensional models to compute entropy  
for both asymptotically flat and AdS black holes. 
 One observes that the resultant entropy follows the Page curve, thanks to the contribution of the 
 island,  despite the fact that the 
 corresponding gravitational models  might  be non-unitary.

\end{abstract}

\end{titlepage}

\section{Introdunction}

According to Hawking's computations the black hole's radiation is thermal \cite{Hawking:1974sw}
which results in black hole information paradox \cite{Hawking:1976ra}. More precisely, being 
thermal, it leads to the fact that the entanglement entropy of the radiation grows monotonically.
This, in turn, implies that the  black hole generates more  entropy than it has room for. 
It is in contrast with the Page's considerations  \cite{{Page:1993wv},{Page:2013dx}}  which propose 
that the corresponding  entropy should decrease after the so called the Page time. 
This is, indeed, what  is required by the  unitarity of quantum mechanics.

In the context of black hole physics it was generally believed that in order to explain the Page 
curve one might need to have a better understanding of microscopic description 
of black hole degrees of freedom. Nonetheless, recently, it was shown that  the Page curve 
could be described within  the semiclassical description of 
 gravity \cite{{Penington:2019npb},{Almheiri:2019psf},{Almheiri:2019hni},{Almheiri:2019yqk}, {Bousso:2019ykv}},
 at least in two dimensions.

Indeed motivated by holographic entanglement entropy \cite{{Ryu:2006bv},{Hubeny:2007xt}}
and introducing the quantum extremal surface \cite{Engelhardt:2014gca}, a new rule
for computing the fine grained black hole entropy is proposed  in \cite{Almheiri:2019hni} based on which 
in order to evaluate the entropy of the radiation one should also consider  a possible contribution 
of an island containing a part of the black hole interior. More precisely, the 
generalized entropy for
a region $R$ is given by \cite{Almheiri:2019hni}
\be\label{IS}
S_{\rm gen}={\rm Min}\;\left\{ {\rm Ext}\left[\frac{A[\partial I]}{4G_N}+S_{\rm vN}(R\cup I)
\right]\right\},
\ee
where $I$ is the island whose boundary area is denoted by $A[\partial I]$ and 
$S_{\rm vN}(R\cup I)$ is the von Neumann entropy of union of the island and the region $R$. 
Then the rule is to extremize this expression for any possible island and then take the one 
that results in the minimum entropy. 

For two dimensional Jackiw-Teitelboim gravity \cite{{Jackiw:1984je},
{Teitelboim:1983ux}} the island rule has been derived by making use of replica trick 
\cite{{Penington:2019kki},{Almheiri:2019qdq}}\footnote{Page curve 
for  evaporating black holes in Jackiw-Teitelboim gravity has also been studied in 
\cite{Hollowood:2020}.}. In this context the island contribution is associated 
with the contribution of new saddle points in the Euclidean path integral (replica wormholes).

It was then wondering if such a prescription  is a particular property of the 
Jackiw-Teitelboim gravity which is conjectured to provide a gravitational description for 
SYK model \cite{{Sachdev:1992fk},{Kitaev}} that enjoys a disorder average procedure. 
Actually, the island rule has been applied for yet another interesting two dimensional 
gravitational model  known as CGHS\cite{Callan:1992rs} that admits two 
dimensional asymptotically flat black holes. It was shown that for this
model the entropy follows the Page curve too \cite{{Gautason:2020tmk},{Anegawa:2020ezn},
{Hartman:2020swn}}.

The existence of the island for higher dimensions has been 
investigated in \cite{Almheiri:2019psy}. More recently, the
 Page curve of  asymptotically flat black hole  for dimensions greater than two has been 
 studied in \cite{Hashimoto:2020cas} (see also \cite{Krishnan:2020}). It is then natural to 
 pose the question whether the island rule could be extended for 
 gravitational models containing higher derivative terms. Indeed, this is the 
 aim of the present article  to  explore this possibility.
 
 To be concrete in this paper we shall consider two different four dimensional 
 gravitational actions containing higher curvature terms. The first model has 
 no cosmological constant so that one has to deal with asymptotically flat black holes, 
 while  in the second one  there is  a negative cosmological constant leading 
 to black hole solutions that are asymptotically AdS. Of course, in this case one has to 
 couple the geometry to a bath where the Hawking radiation may be collected. 
 
 By making use of the extended  island formula we compute the entanglement entropy of the 
 Hawking radiation  and observe that in both cases an island appears at late times resulting in 
 an entropy following the Page curve,   despite the fact that both 
 models might be non-unitary. We will back to this point later.

The paper is organized as follows. In the next section we will present the general procedure 
for evaluating the  island formula for entropy in the presence of higher derivative terms. 
Then we use this formula to compute entanglement entropy for the Hawking radiation for 
asymptotically flat two sided and one sided back holes in sections three and four, respectively. 
In section  five we will address the same question for four dimensional critical gravity where the 
corresponding black hole solutions are asymptotically AdS. The last section is devoted 
to discussions.


\section{Island formula for  entropy  }

In this section we would like to extend the island formula for the  entropy to the cases in which
 the gravitational action contains higher derivative terms. Let us consider the following
total action
\be
 I=I_{\rm gravity}+I_{\rm matter},
 \ee
 where the action of  gravity part may be given by
 \be\label{act}
I_{\rm gravity}=\frac{1}{16\pi G_N}\int d^{d+1} x \sqrt{g}\, \cal{L}(R_{\mu\rho\nu\sigma}) \,.
\ee
 with ${\cal L}$ being a function  constructed  out of contractions of an arbitrary number of 
 Riemann tensors, and  $G_N$ is the Newton constant. Moreover,  $I_{\rm matter}$ stands for 
 the action of quantum matter field propagating on a classic solution of the gravity part. For 
 the  quantum matter  field we may consider the action of $N$ scalar fields.
  In what follows we assume $1\ll N \ll \frac{r_h^{d-1}}{G_N}$, so that the matter  
  contributions dominate the entanglement entropy, while at the same time  the back 
  reaction of the scalar fields on the geometry is negligible.  Here $r_h$ is the horizon 
  radius of  the black hole solution we are considering. 

Since the model under consideration has higher derivate terms, a natural 
proposal for island formula is  to replace the 
area term in the island formula  \eqref{IS} with a proper area 
functional \cite{Dong:2013qoa}. More precisely, for the gravitational  action \eqref{act}  one has
\be\label{IS2}
S_{\rm gen}={\rm Min}\;\left\{ {\rm Ext}\left[S_{\rm gravity}+S_{\rm vN, matter}
\right]\right\},
\ee
with
\bea
&&S_{\rm gravity}= \frac{1}{4G_N}\int_\Sigma d^{d-1} y \sqrt{h} \big\{ -\frac{\partial {\cal L}}
{\partial R_{\mu\rho\nu\sigma}} \varepsilon_{\mu\rho} \varepsilon_{\nu\sigma} + \sum_\alpha 
\left(\frac{\partial^2 {\cal L}}{\partial R_{\mu_1\rho_1\nu_1\sigma_1} \partial 
R_{\mu_2\rho_2\nu_2\sigma_2}}\right)_\alpha \frac{2K_{\lambda_1\rho_1\sigma_1} 
K_{\lambda_2\rho_2\sigma_2}}{q_\alpha+1} \cr &&\cr
&& \;\;\;\;\;\;\;\;\;\;\;\;\;\;
\times
 \left[ (n_{\mu_1\mu_2} n_{\nu_1\nu_2}-\varepsilon_{\mu_1\mu_2} \varepsilon_{\nu_1\nu_2}) 
 n^{\lambda_1\lambda_2} + (n_{\mu_1\mu_2} \varepsilon_{\nu_1\nu_2}+\varepsilon_{\mu_1\mu_2} 
 n_{\nu_1\nu_2}) \varepsilon^{\lambda_1\lambda_2}\right] \bigg\} \,.
\eea
Here in terms of two orthogonal unit vectors $n^{i}_\mu$ to the co-dimension two hypersurface
$\Sigma$, one has
\be
n_{\mu\nu} = n^{i}_\mu n^{i}_\nu g_{ij},\;\;\;\;\;\;\;\;\;\;
\varepsilon_{\mu\nu} = n^{i}_\mu n^{j}_\nu \varepsilon_{ij} \,,
\ee
where $\varepsilon_{is}$ is the usual Levi-Civita tensor.  For more details and 
convention  see \cite{Dong:2013qoa}.

In this paper, to be more concrete, we will  restrict ourselves to four dimensional theories in 
which the corresponding action containing higher derivative terms may be given as follows
\bea\label{action}
I_{\text{gravity}} = \frac{1}{16\pi G_{N}} \int d^{4}x \sqrt{g} \bigg(R[g] + \lambda_{1} R^2[g]+
\lambda_2 R_{\mu\nu}[g]R^{\mu\nu}[g]+\lambda_{\text{GB}}\hspace{.5mm}\mathcal{L}_{GB}[g]
\bigg),
\eea
where
\bea
\mathcal{L}_{\text{GB}}[g] = R_{\mu\nu\rho\sigma}[g]R^{\mu\nu\rho\sigma}[g]-4R_{\mu\nu}
[g]R^{\mu\nu}[g]+R^2[g],
\eea
Indeed  this is the most general four dimensional 
gravitation action consisting of higher derivative terms up to order of $\mathcal{O}(R^3)$. 
The Schwarzschild  black hole solution of the model is given by  
\bea\label{g}
ds^2 = -f(r) dt^2 +\frac{ dr^2}{f(r)} +r^2 d\Omega^2,\hspace{1cm}f(r) = 1-\frac{r_h}{r}
\eea 
that is an asymptotically  flat black hole solution  whose Hawking  temperature is 
$T=\frac{1}{4\pi r_h}$.  On the other hand  by making use of the  Wald formula for the entropy 
one has
\be\label{S}
S_{\rm th}=\frac{\pi}{G_N}(r_h^2+4\lambda_{\rm GB}),
\ee
that is the thermal entropy of the black hole.

For this model the gravity part appearing in the island formula \eqref{IS2} is given by
\cite{Fursaev:2013fta, Dong:2013qoa}.
\be\label{Sgr}
S_{\text{gravity}}
= \frac{A[\partial I]}{4G_{N}} +\frac{1}{4G_{N}} \int_{\partial I}  \left(2\lambda_1 R[g]
+\lambda_2 \left[R_{\mu\nu}[g]n^{\mu}_{i}n^{\nu}_{i}-\frac{1}{2}K_{i}K_{i}\right]
+2\lambda_{\text{GB}}\hspace{.5mm}R[\partial I]\right).
\ee
Here $i=1,2$ denotes two transverse directions to the  co-dimension two boundary of 
island $I$ on which the 
two unit normal vectors are denoted by  $n_{i}^{\mu}$. Moreover, $K_{i}$ is the trace of the 
second fundamental form $K_{i,\mu\nu} =-h_{\mu}^{\alpha}h_{\nu\beta} 
\nabla_{\alpha}n_{i}^{\beta}$ where $h_{\mu\nu}= g_{\mu\nu}-n_{i,\mu}n_{i,\nu}$ 
is the induce metric on $\partial I$. 

As for the contribution of the matter field one needs to compute the von Neumann entropy
  which in four dimensions has the following general form \cite{Solodukhin:2008dh}
\bea\label{Svn}
S_{\text{vN,matter}} = \frac{A[\partial I]}{\epsilon^2} + \tilde{S} \log{\epsilon}+S_{\text{vN,fin}},
\eea
where $S_{\text{vN,fin}}$ is the finite part of the entanglement entropy, $\epsilon$ is a UV cutoff
and 
$\tilde{S} = A \hspace{.5mm}\tilde{S}_{\rm Euler}+
 C\hspace{.5mm} \tilde{S}_{\rm Weyl}$ with
\bea
&&\tilde{S}_{\rm Euler} = \alpha \int_{\partial I}\hspace{1mm} R[{\partial I}],
\cr \nonumber\\
&& \tilde{S}_{\rm Weyl} = -\alpha \int_{\partial I}\hspace{1mm}\left(R_{\mu\nu\alpha\beta}[g]
n^{\mu}_{i}n^{\nu}_{j}n^{\alpha}_{i}n^{\beta}_{j}-R_{\mu\nu}[g]n^{\mu}_{i}n^{\nu}_{i}+\frac{1}{3}
R[g]- \text{Tr}K^2+\frac{1}{2}K_{i}K_{i}\right).
\eea
Here  "$\alpha$" is a finite constant number (see \cite{Solodukhin:2008dh}) and $A , C$ are  the 
coefficients of the Euler term and the Weyl square term in 4D conformal anomaly, respectively. 
These two constants 
play the role of the central charges in four dimensions which are of order of $N$.
On the other hand by  making use the  Gauss-Godazzi equation,
\bea
R[g] = R[{\partial I}] - R_{\mu\nu\alpha\beta}[g]n^{\mu}_{i}n^{\nu}_{j}n^{\alpha}_{i}n^{\beta}_{j}
+2R_{\mu\nu}[g]n^{\mu}_{i}n^{\nu}_{i}+ \text{Tr}K^2 -K_{i}K_{i},
\eea
the von Neumann entropy (\ref{Svn}) associated with the matter field may be  simplified as 
follows 
\bea\label{Svn2}
S_{\text{vN}} = \frac{A[{\partial I}]}{\epsilon^2}\hspace{-1mm}+\alpha \int_{\partial I}\left(\frac{2C}{3}R[g]-C
\left[R_{\mu\nu}[g]n^{\mu}_{i}n^{\nu}_{i}-\frac{1}{2}K_{i}K_{i}\right]+(A-C) R[{\partial I}]\right)
\log{\epsilon}+\hspace{-1mm}S_{\text{vN,fin}}.
\eea
Putting both contributions given by equations  \eqref{Sgr} and \eqref{Svn2} together, it becomes 
clear that the UV divergences of von Neumann  entropy of the matter field may be 
absorbed  by a renomalization of the Newton constant, 
as well as the coupling constants of higher derivative terms. 
More precisely, due to the matter loops the Newton constant will be renormalized as
\[
\frac{1}{4G_N}\rightarrow \frac{1}{4G_N}-\frac{1}{\epsilon^2}\, ,
\]
and this will cancel out the area divergent term in $S_{\text{vN,matter}}$ (see e.g. \cite{Almheiri:2020cfm}). In a similar way the log divergent term in $S_{\text{vN,matter}}$ will be dropped out by the corrections which have root in the renormalization of the coupling constants of the higher derivative terms ($\lambda$'s in \eqref{Sgr}), so although we start with a four dimensional formula for the entropy in its full structure form, we end up with a finite term, $S_{\text{vN,fin}}$ at the end of the day. This term will quantify the mutual information between the regions constructed in radiation part as well as the Island. Since this is very hard to find a universal mutual information in four dimensions we just use the above argument about the propagation of the $s$-modes to estimate the mutual information with a two dimensional formula for which we have a closed form whose universality has already been proved.
Putting things together, one finally arrives at
\bea\label{Sgen}
&&\hspace{-1cm}S_{\text{gen}}(R) = \text{Min}\bigg\{\text{Ext}\bigg[\frac{A[\partial I]}{4 
G_{\text{N,ren}}}
\cr \nonumber\\
&& +\frac{1}{4G_{\text{N,ren}}} \int_{\partial I} \bigg(2\lambda_{1,\text{ren}}R[g] +
\lambda_{2,\text{ren}}\sum_{i=1}^{2}\left[R_{\mu\nu}[g] n^{\mu}_{i}n^{\nu}_{i}-\frac{1}{2}K_{i}K_{i}
\right]+2\lambda_{\text{GB},\text{ren}}\hspace{.1mm}R[\partial I]\bigg)
\cr\nonumber \\
&&\hspace{1.2cm}+S_{\text{vN.fin}}(R \cup I)\bigg]\bigg\}.
\eea
This is, indeed, the island formula we will use to compute the entanglement entropy of 
the Hawking radiation for the four dimensional  asymptotically flat black hole \eqref{g}. 
In fact this should be thought of the semi-classical prescription that computes the fine 
grained entropy of a black hole when higher derivative  terms are also taken into account.


\section{ Entanglement entropy  for two sided black hole}

In this section we would like to compute the entanglement entropy of the Hawking radiation
for an eternal black hole solution given by \eqref{g}  
 using the island formula \eqref{Sgen}. To be concrete we consider subregions
$R_+$ and $R_-$ in the radiation part (see figure 1). In an asymptotically flat black hole
the radiation region is the part of the spacetime near the null infinity where the gravity 
is negligible. Practically,  we assume that it is above a fictitious boundary located at  
$r=r_0$ as shown in  the figure 1.  Typically it is  few time greater than the radius of 
the horizon {\it e.g. } $r_0\sim 3r_h$. Therefore 
with this assumption the radial location of the boundary of our entangling regions denoted by
$b_\pm$ must be  greater  than $3r_h$.
\begin{figure}\label{fig1}
\begin{center}
\includegraphics[width=0.4\linewidth]{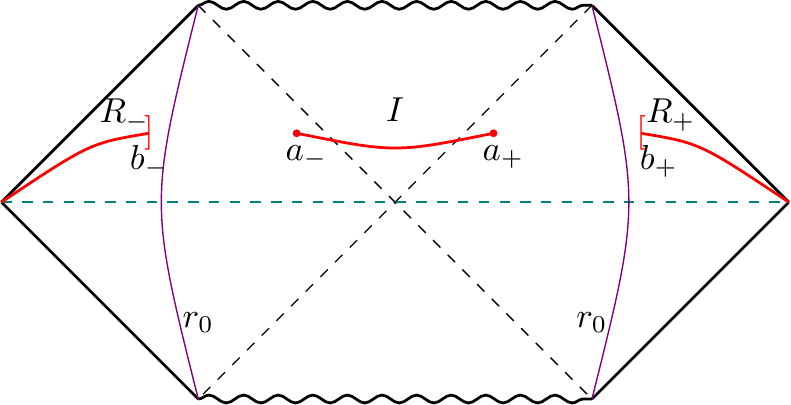}
\caption{Entanglement  regions in the radiation part with the assumption that there is an
island inside the black hole. The fictitious boundaries shown by violet lines at $r=r_0$ are 
the regions over which the gravity is negligible that are the radiation regions.
At early times assuming that there is an island results in  an imaginary  solution for the location of island  indicating that there
is no island at  early times and thus the  whole 
contributions come from the matter von Neumann entropy.  }
\end{center}
\end{figure}

We note that  the island formula consists of two parts: the  gravity part that is 
associated with a nontrivial quantum extremal surface,  island, and 
the matter von Neumann entropy.  A periori, it is not obvious whether or not one should have 
such an island. Nonetheless to proceed in what follows we will
assume that there is a nontrivial island  and then we will seek for its location by 
extremizing the island formula. Note that since the solution we are considering is
maximally symmetric the location of a possible island is fixed by its position in $(t,r)$ coordinates.
Thus, by extremizing the island formula one obtains a set of algebraic equation.

If the resultant algebraic equations have real solution(s), one would get non-trivial island, otherwise one could  conclude that  
there is, indeed, no island and thus the whole contribution to the entropy 
comes from the matter von Neumann entropy.

To proceed,  let us  consider the contribution of the 
gravity part to the entanglement entropy
of the Hawking  radiation assuming that there is an island, $I$,  whose location is denoted
 by $a_\pm$ in figure 1. As we already mentioned, due to the symmetry of the
 solution (\ref{g})  the location of island is give by its position in $(t,r)$ coordinates. Therefore 
 the corresponding  normal vectors $n_{1}$ and $n_{2}$
associated with the co-dimension two boundary of the  island, $\partial I$, are given by 
\bea
n_{1}^{t} = \frac{1}{\sqrt{f(r)}},\hspace{1cm}n_{2}^{r} = \sqrt{f(r)}.
\eea
It is then straightforward to compute  the trace of the extrinsic curvature tensors in two 
normal directions
\bea
K_{1} = 0,\hspace{1cm}K_{2} = -\frac{2}{r}\sqrt{f(r)}.
\eea
We have, now, all ingredients to compute the gravity part of the entropy. Parametrizing  the 
location of the end points of the island by $a_+:(t_a, a)$ and $a_-: (-t_a+i\frac{\beta}{2},a)$ where 
$\beta$ is the inverse of the Hawking temperature, one has 
\bea\label{K}
K_{1}K_{1} =0\ , \hspace{1cm} K_{2}K_{2} = \frac{4}{a^2}\left(1-\frac{r_h}{a}\right).
\eea
Moreover taking into account that for the solution \eqref{g} one has 
$R_{\mu\nu}[g] =0$ and $R[\partial I] = 2/a^2$, the gravitational part of the entanglement entropy of the 
radiation reads
 \bea\label{Sgengenerall}
S_{\text{gravity}} = \frac{2\pi }{G_{\text{N,ren}}}\bigg(a^2-2\lambda_{2,\text{ren}}\left(1-\frac{r_h}{a}\right)+4\lambda_{\text{GB,ren}}\bigg).
\eea

Now one should  compute the matter von Neumann entropy  $S_{\text{vN.fin}}$.
Actually, in general, it is not  an easy task to compute entanglement entropy for several 
intervals in four dimensions. It is, however, worth noting that  we are, actually,  interested in 
evaluating the entanglement entropy of quantum fields on a maximally symmetric 
background containing a 2-sphere. Thus we can expand the quantum fields 
in terms of the spherical harmonics. Reducing to two dimensions 
one gets a tower of Kaluza-Klein modes whose masses are given by the angular moment along 
the 2-sphere.
On the other hand since we are interested in entangling regions that are far farm each other, 
one would expect that the main contribution to the von Neumann  entropy comes from 
entanglement between massless modes; the s-wave configuration 
\cite{{Penington:2019npb},{Hashimoto:2020cas}}. In other words,  from two 
dimensional point of view we are throwing  away the contributions of massive 
Kaluza-Klein modes from the entanglement entropy.
 
In this case, effectively, one might only consider those  modes that propagate in two dimensions
parametrized by $(t,r)$ coordinates. 
As a result we could compute the corresponding entanglement entropy between 
several entangling regions  using two dimensional techniques 
(see for example \cite{Calabrese:2009}). 
To proceed it is useful to work within  the Kruskal coordinates 
\bea\label{UV}
U = - \sqrt{\frac{r-r_h}{r_h}}\hspace{.5mm}e^{-\frac{t-(r-r_h)}{2r_h}},\hspace{.5cm}V = \sqrt{\frac{r-r_h}{r_h}}\hspace{.5mm}e^{\frac{t+(r-r_h)}{2r_h}},
\eea
by which the corresponding two dimensional part of the  metric (\ref{g}) reads
\bea\label{2D}
ds^2 = - \omega^{-2} dU dV,\hspace{1cm}\omega = \sqrt{\frac{r}{4r_h^3}} \hspace{.5mm}
e^{\frac{r-r_h}{2r_h}}\,.
\eea
In this two dimensional theory the finite part  of the entanglement entropy of the regions
$R_+$, $R_-$ and the island $I$ is given by \cite{Calabrese:2009} 
\bea\label{IRpmI}
S_{\rm vN, fin}(R\cup I) = \frac{A}{3}\log\left(\frac{d(a_+,a_-)\hspace{.5mm}d(b_+,b_-)
\hspace{.5mm}d(a_
+,b_+)\hspace{.5mm}d(a_-,b_-)}{d(a_+,b_-)\hspace{.5mm}d(a_-,b_+)}\right),
\eea
where $d(\ell_1,\ell_2)$  denotes the  geodesic length between two-points $\ell_1$ and $\ell_2$
that in the above coordinate system reads
\bea\label{dbpbm}
d(\ell_1,\ell_2)= \sqrt{\frac{\big(U(\ell_2)-U(\ell_1)\big)\big(V(\ell_1)-V(\ell_2)\big)}
{\omega(\ell_1)\omega(\ell_2)}}.
\eea    
To write the above expression for finite part of the entanglement entropy  we have used the
 fact that the whole system represents a pure state and therefore from two dimensional point 
 of view the desired entanglement entropy is the same  as that of two disjoint intervals 
 $[b_-,a_-]\cup [a_+,b_+]$. 

Using this expression the finite part of the entanglement entropy, equation \eqref{IRpmI}
reads \cite{Hashimoto:2020cas}\footnote{Actually this part of computation is almost 
the same as that presented in \cite{Hashimoto:2020cas}.}
\bea
&&S_{\rm vN, fin}(R\cup I) = \frac{A}{6}\log \left[\frac{256 (a-r_h)(b-r_h)r_h^{4}}{a b} 
\cosh^{2}\frac{t_a}{2r_h}\cosh^2\frac{t_b}{2r_h}\right]
\cr\nonumber\\
&&\hspace{3cm}+\frac{A}{3}\log\left[\frac{\frac{1}{2}\sqrt{\frac{a-r_h}{b-r_h}}e^{\frac{a-b}{2r_h}}+
\frac{1}{2}\sqrt{\frac{b-r_h}{a-r_h}}e^{\frac{b-a}{2r_h}}-\cosh\left(\frac{t_a-t_b}{2r_h}\right)}{\frac{1}
{2}\sqrt{\frac{a-r_h}{b-r_h}}e^{\frac{a-b}{2r_h}}+\frac{1}{2}\sqrt{\frac{b-r_h}{a-r_h}}e^{\frac{b-a}
{2r_h}}+\cosh\left(\frac{t_a+t_b}{2r_h}\right)}\right],
\eea
which could be further simplified into the following form assuming that  $a \approx r_h$\footnote{
Actually since we are dealing with large $N$ limit and moreover the entangling regions in the
radiation parts almost cover the whole space, one would expect that the location of 
island to be in the vicinity of the horzion\cite{Almheiri:2019yqk}.} 
\bea\label{IRpmI1}
&&S_{\rm vN, fin}(R\cup I) =\frac{A}{6}\log \left[\frac{256 (a-r_h)(b-r_h)r_h^{4}}{a b} 
\cosh^{2}\frac{t_a}{2r_h}\cosh^2\frac{t_b}{2r_h}\right]
\cr\nonumber\\
&&\hspace{3cm}+\frac{A}{3}\log\left[\frac{\frac{1}{2}\sqrt{\frac{b-r_h}{a-r_h}}e^{\frac{b-a}{2r_h}}-
\cosh\left(\frac{t_a-t_b}{2r_h}\right)}{\frac{1}{2}\sqrt{\frac{b-r_h}{a-r_h}}e^{\frac{b-a}{2r_h}}+\cosh
\left(\frac{t_a+t_b}{2r_h}\right)}\right].
\eea
 Let us first focus on this expression at  early times in which  $T t_a, T t_b \ll 1$. Then, the above 
 expression may be  further simplified as follows
\bea\label{IRpmI2}
&&S_{\rm vN, fin}(R\cup I) = \frac{A}{6}\log \left[\frac{256 (a-r_h)(b-r_h)r_h^{4}}{a b} 
\cosh^{2}\frac{t_a}{2r_h}\cosh^2\frac{t_b}{2r_h}\right]
\cr\nonumber\\
&&\hspace{3cm}-\frac{4A}{3} \frac{\sqrt{a-r_h}}{\sqrt{b-r_h}}\hspace{1mm}e^{\frac{a-b}{2r_h}} 
\cosh\frac{t_a}{2r_h}\cosh\frac{t_b}{2r_h}.
\eea
Plugging the contributions of gravitational part and the above matter part into the 
 generalized entanglement entropy \eqref{Sgen} one  finds
\bea\label{Sgenearly2}
&&\hspace{-1cm}S_{\text{gen}} = \frac{2\pi }{G_{\text{N,ren}}}\bigg(a^2-2\lambda_{2,\text{ren}}
\left(1-\frac{r_h}{a}\right)+4\lambda_{\text{GB,ren}}\bigg)-\frac{4A}{3} \frac{\sqrt{a-r_h}}{\sqrt{b-
r_h}}\hspace{1mm}e^{\frac{a-b}{2r_h}} \cosh\frac{t_a}{2r_h}\cosh\frac{t_b}{2r_h}
\cr\nonumber\\
&&\hspace{1cm} +\frac{A}{6}\log \left[\frac{256 (a-r_h)(b-r_h)r_h^{4}}{a b} \cosh^{2}\frac{t_a}
{2r_h}\cosh^2\frac{t_b}{2r_h}\right],
\eea
that should be extremized with respect to the location of the island, {\it i.e.} with respect to 
$a$ and $t_a$. Actually from the extremization with respect to $a$ one gets 
\bea
a = r_h + \frac{A^{2} G^2_{\text{N,ren}}}{36\pi^2}\frac{ r_h^2 \hspace{.5mm}e^{1-\frac{b}{r_h}}}
{ (b-r_h)(r_h^2-\lambda_2)^2}\cosh^{2}\frac{t_a}{2r_h}\cosh^{2}\frac{t_b}{2r_h}.
\eea
Substituting  back this value of $a$ into (\ref{Sgenearly2}) and then extremizing with respect
 $t_a$ one finds that the resultant equation does not have a real solution indicating that there is
 no island at early times.
 
 As a result, in order to compute the entanglement entropy of the Hawking radiation at 
 early times one only  needs  to consider  the contribution of the von Neumann entropy of the matter field. More explicitly, assuming there is no island one  has
 \bea\label{Mut2}
S_{\rm gen}=S_{\rm vN, fin}(R_+\cup R_-) = \frac{A}{3} \log d(b_+,b_-)=\frac{A}{6} \log \frac{\big(U(b_-)-U(b_+)\big)\big(V(b_+)-V(b_-))}{W(b_+)W(b_-\big)},
\eea
which yields
\bea\label{Sgenearly}
S_{\text{gen}} = \frac{A}{6} \log \left(16 r_h^2\hspace{.5mm} (\frac{b-r_h}{b})\cosh^2 \frac{t_b}{2r_h}\right).
\eea
Note that to write the above expression for entropy we have again used the fact that
the whole system is in a pure state and to find the desired entropy one just need to evaluate 
the entanglement entropy of a signal interval $[b_-,b_+]$ in  two dimensions. Form this
expression one finds that at  early times it exhibits  $\mathcal{O}(t^{2})$  growth, while at the 
late times  the entropy grows linearly 
\bea\label{Slate51}
S_{\text{gen}} \sim \frac{A}{6}\frac{t_b}{r_h},
\eea
that leads to information paradox as proposed by Hawking.  It is important to note that the
 above computations were based on an assumption that the no-island scenario remains valid 
 all the way  from early to  late times. As we will see this is, actually, not the case.

To explore this point better let us redo the same extrimazation procedure when one is 
approaching the late times. In this
case, assuming $T t_b, T t_a \gg 1$, the equation \eqref{IRpmI1} reduces to
\bea\label{Svnfinout}
S_{\rm vN, fin}(R\cup I) = \frac{A}{6}\log\left(16 r_h^{4}\hspace{.5mm}\frac{(b-r_h)^2
\hspace{.5mm} e^{\frac{b-a}{r_h}}}{a b}\right)-\frac{2A}{3}\sqrt{\frac{a-r_h}{b-r_h}}
\hspace{.5mm}e^{\frac{a-
b}{2r_h}}\cosh\left(\frac{t_a-t_b}{2r_h}\right).
\eea
Therefore, in this case the generalized entropy reads
\bea\label{Sgenout1}
S_{\text{gen}} &=& \frac{2\pi }{G_{\text{N,ren}}}\bigg(a^2-2\lambda_{2,\text{ren}}\left(1-\frac{r_h}
{a}\right)+4\lambda_{\text{GB,ren}}\bigg)+ \frac{A}{6}\log\left(16 r_h^{4}\hspace{.5mm}\frac{(b-
r_h)^2\hspace{.5mm} e^{\frac{b-a}{r_h}}}{a b}\right)\cr &&\cr
&&-\frac{2A}{3}\sqrt{\frac{a-r_h}{b-r_h}}\hspace{.5mm}e^{\frac{a-b}{2r_h}}\cosh\left(\frac{t_a-t_b}
{2r_h}\right).
\eea
From the exterimazation of the generalized entropy one gets 
\bea\label{aout}
\frac{\partial S_{\text{gen}}}{\partial a} = -\frac{A\hspace{.5mm} e^{\frac{r_h-b}{2r_h}}}{3\sqrt{b-
r_h}}\cosh\left(\frac{t_a-t_b}{2r_h}\right)\frac{1}{\sqrt{a-r_h}}+\frac{12\pi (r_h^2-\lambda_2)-A
\hspace{.5mm} G_{\text{N,ren}}}{3r_h G_{\text{N,ren}}}+\mathcal{O}(\sqrt{a-r_h})= 0,
\eea
that may be solved to find
\bea\label{aout2}
a = r_h + \frac{A^2 r_{h}^2\hspace{.5mm}G_{\text{N,ren}}^2 }{144\pi^2(r_h^2-
\lambda_2)^2}\frac{e^{\frac{r_h-b}{r_h}}}{(b-r_h)}\cosh^2\left(\frac{t_a-t_b}{2r_h}\right).
\eea
Here we have used the fact that at leading order one has $12\pi (r_h^2-\lambda_2)-A
\hspace{.5mm} G_{\text{N,ren}} \approx 12\pi (r_h^2-\lambda_2)$. Substituting the above value of $a$ in (\ref{Sgenout1}) and extremizing the result with respect to 
$t_a$ one finds that  $t_a = t_b$ is a  solution. 
Thus, one gets nontrivial real solution for the parameters of island indicating 
that the island shows up at late times when $T t_b\gg  1$. Actually for this solution
 the generalized entropy (\ref{Sgenout1})  at late times reads
\bea\label{Sgenout2}
S_{\text{gen}} = 
2S_{\text{th}}+\frac{A}{6}\log\left(16 r_h^3\hspace{.5mm} \frac{(b-r_h)^2}{b}\hspace{.5mm}e^{\frac{(b-r_h)}{r_h}}\right) +\mathcal{O}(G_{\text{N,ren}}),
\eea
where $S_{\rm th}$ is black hole thermal entropy \eqref{S}.
It is also  worth noting that the  contribution of $\lambda_2$ term appears in order
 $G^{2}_{\text{N,ren}}$. From this result it is then clear that the appearance of the 
 island  results in the saturation of entanglement entropy at late times in agreement 
 with Page's proposal.  

It is also possible to estimate the time when the entropy stops growing that is known as the Page time. Indeed, this can be done by equating the entropy growth without island
\eqref{Sgenearly} with the saturation value at late times. Doing so, one arrives at
\be
  \log \left(\cosh \frac{t_{\rm Page}}{2r_h}\right)
 = \frac{6S_{\text{th}}}{A}+\frac{1}{2}\log\left(r_h(b-r_h)\hspace{.5mm}e^{\frac{(b-r_h)}{r_h}}\right). 
 \ee
Tending to the late time limit $T t_b\gg 1$, the 
above equation may be simplified as follows
\be  
t_{\rm Page}
 = \frac{12S_{\text{th}}}{A}\, r_h+r_h\,\log\left(r_h(b-r_h)\hspace{.5mm}e^{\frac{(b-r_h)}{r_h}}\right). 
\ee

\section{Entanglement entropy for one sided black hole}

In this section we would like to study  entanglement entropy of the Hawking radiation 
of an asymptotically flat one sided black hole when the corresponding gravitational action
contains higher curvature terms. We will consider an entangling 
region, $R$, in the radiation part of the black hole  that is the part near null infinity  
behind a fictitious  boundary  over which the gravity is negligible (see  figure 2). 
\begin{figure}\label{fig2}
\begin{center}
\includegraphics[width=0.24\linewidth]{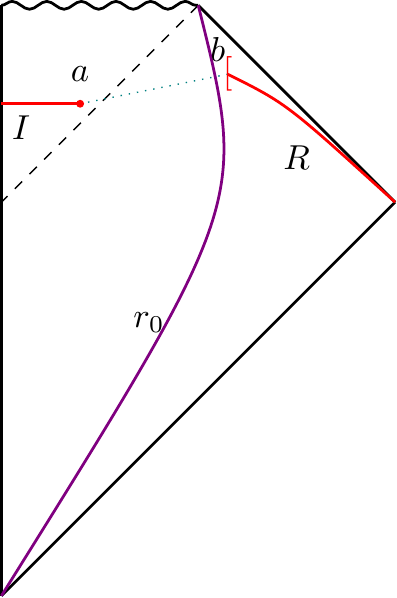}
\caption{Entanglement  regions in the radiation part with the assumption that there is an
island inside the black hole. The fictitious boundary shown by violet lines  at $r=r_0$ specifies 
the region over which the gravity is negligible, {\it{i.e.}} the radiation region. }
\end{center}
\end{figure}

Actually  the aim  is to evaluate the island formula for this configuration. 
To do so, one needs to compute the generalized entropy assuming that there is 
an island whose location can be obtained by extreminzing the generalized entropy \eqref{Sgen}.
It is important to note that while in two sided black hole the position of the island was outside 
the horizon, in the present case it is believed that the island is located inside 
the event 
horizon \cite{{Penington:2019npb},{Almheiri:2019psf},{Almheiri:2019hni},{Almheiri:2019yqk}}. 
Having this point in mind the gravity contribution to the generalized entropy is found to be
\bea\label{onegr}
S_{\text{gravity}}=\frac{\pi }{G_{\text{N,ren}}}\bigg(a^2+2\lambda_{2,\text{ren}}\left(1-\frac{r_h}{a}
\right)+4\lambda_{\text{GB,ren}}\bigg).
\eea

To find the von Neumann entropy of the matter part, following the procedure we used in the 
previous section, one may consider entanglement entropy of a two dimensional space 
given by the metric \eqref{2D}. Of course, since the black hole and radiation union
is in a pure state, in order to compute the entanglement entropy of radiation and island one 
just needs to compute the entanglement  entropy of an interval $[a, b]$ in two dimensions that is
given by 
\be
S_{\rm vN, fin}=\frac{A}{3}\log d(a,b)
\ee
where $d(a,b)$ is the geodesic distance between $a$ and $b$ as depicted in the figure 2.
To compute the corresponding distance one should use the equation \eqref{dbpbm}. 
It is, however, important to note that since the island is located 
behind the horizon one should properly define the Kruskal coordinates in this case. More 
precisely one has
\bea\label{UV1}
&&U =  - e^{-\frac{t-r^{*}}{2r_h}},
\hspace{1cm} V=  e^{\frac{t+r^{*}}{2r_h}}, \hspace{1cm}\text{outside the horizon},
\cr \nonumber\\
&&U =  e^{-\frac{t-r^{*}}{2r_h}},
\hspace{1cm} V = e^{\frac{t+r^{*}}{2r_h}},\hspace{1.4cm}\text{inside the horizon},
\eea
where the tortoise coordinate, $r^*$, is given by
\bea\label{rstar}
r^{*} = r-r_h+r_h \log\left(\frac{\mid \hspace{-.5mm}r-r_h\hspace{-.5mm}\mid}{r_h}\right).
\eea
By making use of this notation and utilizing the equation \eqref{dbpbm} one finds 

\bea\label{onevN}
S_{\text{vN,fin}} &=&\frac{A}{6}\log\bigg[(a-r_h)e^{\frac{a-b}{2r_h}}+(b-r_h)e^{\frac{b-a}{2r_h}}-2\sqrt{(b-r_h)(r_h-a)}\sinh\left(\frac{t_a-t_b}{2r_h}\right)\bigg]
\cr \nonumber\\
&&+\frac{A}{6}\log\bigg[\frac{4r_h^2}{\sqrt{ab}}\bigg]
\eea

To find the location of the  island and then the  entanglement entropy of the radiation one  
needs to extremize the generalized entropy with respect to the location of the island.
Indeed extremizing the generalized entropy with respect to $a$,
\bea\label{diffa}
\frac{\partial S_{\text{gen}}}{\partial a} =\frac{\partial}{\partial a}\left( S_{\text{gravity}}+S_{\text{vN,fin}}\right) =0,
\eea
and  defining  $X = \sqrt{\frac{r_h-a}{r_h}}$  one arrives at 
\bea\label{rr}
&&\frac{12\pi (r_h^2+\lambda_{2,\text{ren}})-A G_{\text{N,ren}}}{G_{\text{N,ren}}} + \frac{r_h A}{(b-
r_h)}\left( e^{\frac{r_h-b}{r_h}} \cosh\left(\frac{t_a-t_b}{r_h}\right)+ e^{\frac{r_h-b}{2r_h}}
\sqrt{\frac{b-r_h}{r_h}}\hspace{1mm}\frac{\sinh\left(\frac{t_a-t_b}{2r_h}\right)}{X}\right)
\cr \nonumber\\
&&\hspace{2cm}+ \mathcal{O} \bigg(X\sinh\left(\frac{t_a-t_b}{2r_h}\right)\bigg) = 0.
\eea
It is easy to see that at early times when $t_a \sim t_b $ and $Tt_b\ll 1$ the above equation has no 
solution and therefore one may conclude that at early times there is no island. Thus the whole 
contribution to the generalized entropy comes from the matter von Neumann entropy. More 
precisely, in this case one gets 
\bea
S_{\text{gen}}  =\frac{A}{6}\log\bigg[\frac{4r_h^2}{\sqrt{b}}\bigg((b-r_h)e^{\frac{b}{2r_h}}-r_h
e^{\frac{-b}{2r_h}}+2\sqrt{(b-r_h)r_h}\sinh(\frac{t_b}{2r_h})\bigg)\bigg],
\eea
that results in the following linear growth at early times 
\bea
S_{\text{gen}} \sim \frac{A}{6} \frac{\sqrt{(b-r_h)r_h}}{r_h\left((b-r_h)e^{\frac{b}{2r_h}}-r_h
e^{\frac{-b}{2r_h}}\right)}\hspace{1mm}t_b\,.
\eea
Assuming to have no island all the time from early to late times one observes that 
the entropy increases monotonically, that is consistent with Hawking's proposal. Of course 
this is not the case as we will see.

Indeed at late times when $T t_b\gg1 $  the equation \eqref{rr} admits a  solution 
that is given by 
\bea\label{X}
&&\hspace{-.7cm}X = - \frac{A\hspace{.5mm}G_{\text{N,ren}}\hspace{1mm} e^{\frac{r_h-b}{2r_h}}
\hspace{.5mm}  \hspace{1mm}\sqrt{r_h\big(b-r_h\big)}\sinh\left(\frac{t_a-t_b}{2r_h}\right)}
{\bigg(\big(b-r_h\big)\big(12\pi r_h^2 +12\pi\lambda_{2,\text{ren}}-A G_{\text{N,ren}}\big)+A 
G_{\text{N,ren}} \hspace{.5mm}e^{1-\frac{b}{r_h}} r_h\cosh\left(\frac{t_a-t_b}{r_h}\right)\bigg)}.
\eea
Note that since by definition the quantity $X$ is a positive number, the above expression leads 
to a solution if $t_a < t_b$. In other words the location of the island should be in the past of 
the location of the entangling region $R$. Moreover, from the extremization with 
respect to $t_a$ one finds that\footnote {Here we have used 
the approximation   $12\pi r_h^2 +12\pi\lambda_{2,\text{ren}}-A G_{\text{N,ren}} \approx 12\pi 
r_h^2 +12\pi\lambda_{2,\text{ren}}$.}
\bea
&&\hspace{-1cm}\frac{\partial S_{\text{gen}}}{\partial t_a} = \frac{A\hspace{.5mm}G_{\text{N,ren}}
\hspace{1mm}e^{\frac{b}{r_h}}}{(b-r_h)(r_h^2+\lambda_{2,\text{ren}})}\hspace{.5mm}\sinh
\left(\frac{t_a-t_b}{r_h}\right) 
\\
&&\hspace{-.7cm} \times\left(\big(b-r_h\big)\big(A G_{\text{N,ren}}+12\pi(r_h^2+
\lambda_{2,\text{ren}})\big)+A\hspace{.5mm}G_{\text{N,ren}} e^{1-\frac{b}{r_h}}r_h\big(1-2\cosh
\left(\frac{t_a-t_b}{r_h}\right)\big)\right) = 0,\nonumber
\eea
which can be solved to find (note that $t_a< t_b$)
\bea\label{ta}
t_a = t_b - r_h\log S_{\text{th}} -r_h\log\left(\frac{12\hspace{.5mm}e^{\frac{b}{r_h}-1}(b-r_h)(r_h^2+\lambda_{2,\text{ren}})}{A\hspace{.5mm}r_h (r_h^2+\lambda_{\text{GB,ren}})}\right)+\mathcal{O}(G_{\text{N,ren}}).
\eea
By making use of the definition  of the tortoise coordinates the above equation may be 
recast into the following form
\bea
v(a,t_a) - v(b,t_b) \sim -r_h \log S_{\text{th}} +\mathcal{O}(G^{0}_{\text{N,ren}}) \sim -\frac{
 t_{\text{scr}}}{2},
\eea
where  $v=t+r^*$ and $t_{\text{scr}}$ is the scrambling time. This is, indeed, a realization 
of  the Hayden-Preskill decoding criterion \cite{Hayden:2007}. Namely, if one throws a quantum
q-bit into the black hole after the Page time, it can be decoded from the Hawking radiation just after 
waiting for a time of order of the scrambling time.

Finally using this result one may find the generalized entropy at late times when both 
contributions of island and matter field should be taken into account 
\bea
S_{\text{gen}} = S_{\text{th}}-\frac{A}{12}\left(1-\log\bigg[\frac{16 e^{\frac{b}{r_h}}(b-r_h)^2 r_h^3}{b}\bigg]\right)+\mathcal{O}(G_{\text{N,ren}}).
\eea

To conclude this section we note that although at early times one has a linear growth, the island 
comes to rescue the unitarity at late times in agreement with the Page curve. In the
present case the Page time at leading order is 
\be
t_{\rm Page}\sim \frac{6 S_{\rm th}}{A} r_h.
\ee


\section{Page curve for critical gravity }

In this section we will explore the behavior of the  Page curve  for the black hole solutions of  
yet another interesting gravitational  theory containing higher derivative terms.  
The model we will be considering  is ``critical gravity''  of which the  action
in four dimensions is  \cite{Lu:2011zk}
 \bea\label{HD}
I_{\rm critical} = \frac{1}{16\pi G}\int_{\mathcal{M}} d^{4}x \sqrt{-g}\left[R-2\Lambda-\frac{1}{m^2}
\left(R^{\mu\nu}R_{\mu\nu}-\frac{1}{3}R^{2}\right)\right],
\eea
where $m$ is a dimensionful parameter. This model admits several solutions including
AdS and AdS black holes with radius $\ell^2=-\frac{3}{\Lambda}$. It is known that at the critical point
where  $m^2=\frac{8}{\ell^2}$ the model degenerates yielding to a log-gravity
\cite{Alishahiha:2011yb}. Of course in what follows we will study the model away from 
the critical point, {\it i.e.} $m^2\neq\frac{8}{\ell^2}$. 

It is easy to see that the equations of motion obtained from the action \eqref{HD}
admits the AdS-Schwarzschild black hole whose metric may be written as follows
\be\label{metric}
ds^{2} =\frac{dr^{2}}{f(r)}-f(r)dt^{2}+r^{2}d\Omega^{2}_{2},\;\;\;\;\;\;\;\;\;\;
f(r) = \frac{r^{2}}{\ell^{2}}+1-\frac{4GM}{r}.
\ee
where $M$ is a parameter of the solution (not the physical mass).

For our purpose one needs to couple the above gravitational model to a quantum field that propagate in the above geometry.
It is important to note that unlike the solutions we have considered in the previous section, the 
above metric represents  a geometry that is  asymptotically AdS. The  main difference between 
 asymptotically AdS and asymptotically flat black holes is that, in the former case one
  usually has reflecting boundary condition on bulk fields at the boundary of spacetime. 
 This will eventually 
 cause the termination of the black hole evaporation due to the 
 equilibrium between the emission and the absorption processes.  

To overcome this problem and to have a fully evaporating system,
 following \cite{{Penington:2019npb},{Almheiri:2019psf}}, one should impose transparent 
 boundary  condition for the matter field while the boundary condition for the gravitational field 
 remains unchanged. In other words, one should couple the gravitational theory to an external
 bath constructed of the same quantum matter field propagating in the flat space. 
 So that the bath is a flat spacetime with no gravity where the Hawking radiation 
 could be collected. Therefore altogether we have a system consisting of gravity confined in the 
 AdS geometry and a quantum field that propagate  both in AdS and flat spaces with transparent 
 boundary condition at the boundary of the AdS geometry.  For quantum matter field one may take 
 the action of $N$ scalar fields, see figure 3.
 \begin{figure}\label{fig3}
\begin{center}
\includegraphics[width=0.4\linewidth]{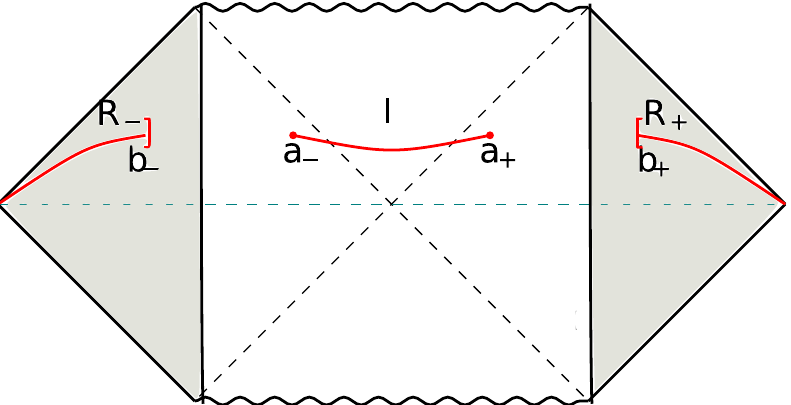}
\caption{The AdS black hole+Bath system. The bath (colored region) plays the role of the environment around the evaporating black hole. This is necessary to include this environment since the boundary of AdS is reflecting and as a result of that the evaporation will be terminated at some point due to the expected equilibrium between the emission and the absorption processes.}
\end{center}
\end{figure}

Here the main subtlety  is the way the boundary conditions should be imposed so that 
the whole system (gravity+bath) becomes a consistent model. The corresponding 
procedure for two dimensional Jackiw-Teitelboim gravity  has been carefully worked 
out in the literature (see {\it e.eg} \cite{{Penington:2019npb},{Almheiri:2019psf},{Almheiri:2019hni},
{Almheiri:2019yqk}}), though its generalization to higher dimensions has not been fully 
studied yet.  Nonetheless in general one would expect that the similar analysis could be 
done for higher dimensions too.
  
As far as the gravitational part of the island formula is concerned, it is 
straightforward to evaluate the corresponding contribution in arbitrary dimensions, though 
for the matter field it is crucial to have the proper conditions under which the bath is connected 
to the asymptotic AdS boundary. Motivated by the results of the previous section 
we note that  since the main contribution to the entanglement entropy  comes from the s-wave 
modes of the quantum field, one may effectively work within a two dimensional theory 
obtained by dimensional reduction from the original 
four dimensional metric. Therefore one would expect that the same boundary conditions as
those in two dimensions may be used here.  Actually this is the fact we assume in what 
follows and accordingly we will use the same  procedure as before  to compute the finite 
part of the entanglement entropy.

For a technical simplicity reason in what follows we will consider the case where the 
radius of curvature  is much larger than the radius of horizon so that we could essentially 
work with a black brane solution whose metric is the same as that in (\ref{metric}) with
\be
f(r) = \frac{r^2}{\ell^2}\left(1-\frac{r_h^3}{r^3}\right).
\ee
It is then easy to evaluate the contribution of gravitational part to the island formula. Actually 
by making use of the fact that in the present case one has \footnote{We are using the
same notation as that in the section three. See also figure 3.}
\bea
K^1 = 0,\hspace{1cm}K^2 = \frac{2}{r}\sqrt{f(r)},\hspace{1cm} R[g] = -\frac{12}
{\ell^2},\hspace{1cm} \sum_{i} R_{\mu\nu}n_i^\mu n_i^\nu =0,
\eea
the gravitational part reads
\bea\label{CGgravity} 
S_{\text{gravity}} = \frac{V_2}{2G_{\text{N,ren}}}\hspace{1mm}\frac{a^2}{\ell^2}\left(1-\frac{8}
{\ell^2m^2}+\frac{2}{\ell^2m^2}(1-\frac{r_h^3}{a^3})\right),
\eea
where $V_2$ is the regularized volume of two dimensional  transverse space. As for the matter part, using the Kruskal coordinate
\bea
U = -e^{-\frac{\beta}{2\pi}(t-r^{*})},\hspace{1cm}V = e^{\frac{\beta}{2\pi}(t+r^{*})}
\eea
the corresponding reduced two dimensional  metric is
\bea
ds^2 = -\omega^{-2}{dU\hspace{.5mm}dV},\hspace{1.5cm}
\omega = \left(\frac{4\pi^2 e^{\frac{4\pi r^{*}}{\beta}}}{\beta^2 f(r)}\right)^{\frac{1}{2}},
\eea
where $\beta=\frac{4\pi}{f'(r_h)}$ is the inverse of the Hawking temperature and $r^{*} = \int dr/f(r)$
is the tortoise coordinate. Then the finite part of the entanglement entropy will be still 
obtained from the equation \eqref{IRpmI}.
It is, however, important to note that in order to compute the geodesic distances one should 
keep in mind that the whole system (BH+bath) should parametrized with the same coordinate
system as above, thought in the bath one should set $f(r) =1$. This technical procedure
guarantees that we are dealing with a  state that is the Minkowski vacuum in the whole system.

Going through the same computations as those in the section three one finds that at
early times, $T t_b\ll 1$,  there is no island and  the whole contribution to the generalized 
entropy comes from the matter part. Assuming that there is no island at all, the corresponding  
generalized entropy becomes
\bea\label{noislCG}
S_{\text{gen}} = \frac{A}{3}\log{d(b_-,b_+)} = \frac{A}{6}\log\left(\frac{\beta^2}{\pi^2}\cosh^{2}\left(\frac{2\pi t_b}{\beta}\right)\right),
\eea
that at early times results in $S_{\rm gen}\sim {\rm const.} +\frac{2\pi^2 A}{3\beta^2} 
\hspace{.5mm}t_b^2$, whereas at late times where it keeps growing linearly as follows
\bea
S_{\text{gen}} \sim \frac{2\pi A}{3\beta}\hspace{1mm} t_b. 
\eea

On the other hand assuming to have an island at  late times one would have to extremize
the generalized entropy  when both the gravitational part and matter entanglement entropy are 
taken into count. In this case using the equation  \eqref{IRpmI}
one arrives at
\bea
&&S_{\text{vN,fin}} =  -\frac{4\pi A}{3\beta}\big(t_a+t_b\big)+\frac{A}{3}\log\left(\frac{\beta^2}
{4\pi^2}\sqrt{f(a)}\hspace{1mm}(1+e^{\frac{4\pi t_a}{\beta}})(1+e^{\frac{4\pi t_b}{\beta}})\right)
\cr \nonumber\\
&&\hspace{1cm}-\frac{A}{3}\log\left(\frac{ \bigg(e^{\frac{2\pi}{\beta}\big(t_a+r^{*}(a)\big)}
+e^{\frac{2\pi}{\beta}\big(-t_b+b\big)}\bigg)\bigg(e^{\frac{2\pi}{\beta}\big(-t_a+r^{*}(a)\big)}
+e^{\frac{2\pi}{\beta}\big(t_b+b\big)}\bigg)}{\bigg(e^{\frac{2\pi}{\beta}\big(t_b+r^{*}(a)\big)}-
e^{\frac{2\pi}{\beta}\big(t_a+b\big)}\bigg)\bigg(e^{\frac{2\pi}{\beta}\big(t_a+r^{*}(a)\big)}-
e^{\frac{2\pi}{\beta}\big(t_b+b\big)}\bigg)}\right),
\eea
which at late times, $t_a,t_b \gg \beta$, reads
\bea
&&S_{\text{vN,fin}} = -\frac{2\pi A}{3\beta}\bigg(t_a +t_b+ r^{*}(a)+b\bigg)+\frac{A}{3}\log
\left(\frac{\beta^2 \sqrt{f(a)}}{4\pi^2}\right)
\cr \nonumber\\
&& +\frac{A}{3} \log\bigg(\big(e^{\frac{2\pi}{\beta}\big(t_b+b\big)}-e^{\frac{2\pi}{\beta}\big(t_a
+r^{*}(a)\big)}\big)\big(e^{\frac{2\pi}{\beta}\big(t_a+b\big)}-e^{\frac{2\pi}{\beta}\big(t_b+r^{*}(a)
\big)}\big)\bigg).
\eea
Therefore one has to extremize the following expression of the generalized entropy 
\bea
S_{\rm gen} &=&  \frac{V_2}{2G_{\text{N,ren}}}\hspace{1mm}\frac{a^2}{\ell^2}\left(1-\frac{8}
{\ell^2m^2}+\frac{2}{\ell^2m^2}(1-\frac{r_h^3}{a^3})\right)\hspace{1mm}
-\frac{2\pi A}{3\beta}\bigg(t_a +t_b+ r^{*}(a)+b\bigg)
\cr \nonumber\\
&&\hspace{-1.5cm}+\frac{A}{3}\log\left(\frac{\beta^2 \sqrt{f(a)}}{4\pi^2}\right)
+\frac{A}{3} \log\bigg(\big(e^{\frac{2\pi}{\beta}\big(t_b+b\big)}-e^{\frac{2\pi}{\beta}\big(t_a+r^{*}
(a)\big)}\big)\big(e^{\frac{2\pi}{\beta}\big(t_a+b\big)}-e^{\frac{2\pi}{\beta}\big(t_b+r^{*}(a)\big)}
\big)\bigg).
\eea
Actually It is straightforward to show that at late times the equation  
\bea
\frac{\partial S_{\text{gen}}}{\partial t_a} = 0,
\eea
implies $t_a =t_b$, while from the equation 
\bea
\frac{\partial S_{\text{gen}}}{\partial  a} =0,
\eea
one finds 
\bea
a = r_h + \frac{ A^2 \ell^8 m^4 G^2_{\text{N,ren}} }{9\sqrt{3}\hspace{.5mm}r_h^3 
\left(\ell^2m^2-5\right)^2 V_2^2}\hspace{1mm}e^{\frac{\pi}{\sqrt{3}}-\frac{3 b\hspace{.5mm} r_h}
{l^2}}.
\eea
Plugging this result into the expression of generalized entropy one arrives at
\bea
S_{\text{gen}} = \frac{V_2 r_h^2}{2\ell^2 G_{\text{N,ren}}}\left(1-\frac{8}{\ell^2m^2}\right)+ \frac{A}{6}\left(\frac{3b\hspace{.5mm} r_h}{\ell^2}-\frac{\pi}{\sqrt{3}}+\log(\frac{16 \ell^6}{9\sqrt{3}r_h^2})\right)+\mathcal{O}(G_{\text{N,ren}}).
\eea
that is 
\bea
S_{\text{gen}} = 2 S_{\text{th}}+ \frac{A}{6}\left(\frac{3b r_h}{\ell^2}-\frac{\pi}{\sqrt{3}}+\log(\frac{16 \ell^6}{9\sqrt{3}r_h^2})\right)+\mathcal{O}(G_{\text{N,ren}}).
\eea
Therefore we find the Page curve for the fine grained entropy for the black brane solution 
in the critical gravity, despite the fact the model is believed to be non-unitary.  Note that 
in this case the Page time is also given by $t_{\rm Page}\sim \frac{12 S_{\rm th}}{A} r_h$.


\section{Discussions}

In this paper we have extended the island formula to  general gravitational theories 
containing higher derivative terms in  diverse dimensions.  Although we could have done
our explicit computations in general dimensions, in order to be concrete,  we have 
restricted ourselves to four dimensional theories with curvature squared terms with and 
without negative cosmological constant. 

For the model without cosmological constant we have evaluated entanglement entropy
of the Hawking radiation for both two sided and one sided asymptotically flat black holes. Whereas
for the case with the negative cosmological constant we have only considered the two sided
black holes that are asymptotically AdS. Although for the asymptotically flat case there was 
a natural region to collect the Hawking radiation, for the asymptotically AdS case we 
had to couple the system to a bath. 

It both cases, under certain reasonable assumptions, we have found that the generalized 
entropy follows the Page curve, despite the fact that both model are non-unitary.  The Page
curve appears due to  the non-trivial contribution from island.

It is important to mention that our results rely on the certain assumptions.  For asymptotically 
flat black holes we have assumed that there is a fictitious surface
over which the gravity is negligible and essentially we have Hawking radiation with no
gravitational interaction. For asymptotically AdS black hole we have assumed that 
the geometry can be consistently coupled to a flat bath with the transparent boundary 
condition on the quantum matter. 

On top of it we have assumed that the main contribution to the matter von Neumann entropy 
comes from the entanglement between s-wave modes of the quantum field that has no
dependence on the 2-sphere. Therefore one might reduce the theory into two 
dimensions in which one could use two dimensional conformal field theory techniques to compute 
the finite term of the corresponding  entanglement entropy. Indeed since we are dealing with 
a spherical symmetric geometry the classical maximin surface (and the eventual
quantum extremal surface) should be rotationally symmetric.  On the other hand each
angular momentum mode acts as an
independent free field in an effective two-dimensional theory, with a Kaluza-Klein mass
proportional to inverse of the radius of sphere  which can be ignored at the lengthscales of interest. More  precisely in order to  rely on this assumption one has to 
consider cases where the geodesic distances between different entangling intervals are grater
than the correlation  length of the massive Kaluza-Klein modes ( see\cite{Penington:2019npb}
for discussions on this point).

An intuitive argument supporting the above assumption may be given by 
studying the scattering amplitude of a scalar field off the black hole. 
Actually, one can see that due to the presence of the black hole each
angular momentum mode which acts as an
independent free field  feels  a repulsive potential given by\cite{Susskind:2005}
\be
V_\ell(r^*)=\frac{r-r_h}{r}\left(\frac{\ell(\ell-1)}{r^2}+\frac{r_h}{r^3}\right),
\ee
where $r^*=r+r_h\log(r-r_h)$ and $\ell$ is the angular momentum along the 
sphere. For $r\gg r_h$ the potential is repulsive and when one is  very close to the 
horizon  the potential is attractive that pulls a wave packet toward the horizon.
 The height of potential  depends on the angular momentum along
 the sphere and the lowest  height is  associated with the s-wave. More precisely, one has \cite{Susskind:2005}
  \be
 V_0^{\rm max}\approx 0.1\;\frac{1}{r_h^2},\;\;\;\;\;\;\;\; V_\ell^{\rm max}\approx 0.15\;
 \frac{\ell^2}{r_h^2}
 \ee
 for large $\ell$. Therefore  while the s-wave modes could 
propagate almost freely the higher modes feel a barrier. Therefore it is natural to assume that
the main contribution to the entanglement entropy comes from the s-wave. 

Note also that  assuming  the fact that the s-wave is responsible for generating 
entanglement entropy the desired boundary condition to attach the asymptotically AdS 
geometry to a flat bath may reduce to that of the two dimensional theory. 

With these conditions we have shown that the proposed island formula for entropy 
offers an expression for the  fine grained entropy that follows the Page curve. It is important 
to note that this result has nothing to do with holography and having obtained the Page 
curve is a general feature of any gravitational theory, despite the fact the island formula is
originally motivated in the context of AdS/CFT correspondence.

Actually  another way to think of the island formula is that it is the correct formula to compute
fine grained entropy of a black hole  imposing the Page curve as a guiding principle. 
In fact this is the point Hawking missed in his original computations of radiation entanglement 
entropy. Indeed  this is reminiscent of coarse grained 
entropy of a black hole  in which imposing to have 
the second law of thermodynamics leads us to define generalized entropy for black holes.

Thinking in this way, it is then evident (as it is anticipated in the literature)  that obtaining 
the Page curve is not the full story to resolve the black hole information paradox. It is just
one step forward to make it precise how to compute entanglement entropy for a system 
involving gravitational interaction. In fact full resolution of information paradox requires
better understanding of the quantum state of  Hawking radiation and the dynamics of the system.

\section*{Acknowledgments}

We would like to thank Mehregan Doroudiani, G.  Jafari, A. Mollabashi, M. R. Mohammadi Mozaffar and B. Taghavi for useful discussions.

\end{document}